# Benchmarking Data Analysis and Machine Learning Applications on the Intel KNL Many-Core Processor

Chansup Byun, Jeremy Kepner, William Arcand, David Bestor, Bill Bergeron, Vijay Gadepally, Michael Houle, Matthew Hubbell, Michael Jones, Anna Klein, Peter Michaleas, Lauren Milechin, Julie Mullen, Andrew Prout, Antonio Rosa, Siddharth Samsi, Charles Yee, Albert Reuther

MIT Lincoln Laboratory Supercomputing Center, Lexington, MA, U.S.A

*Abstract*— **Knights Landing (KNL) is the code name for the second-generation Intel Xeon Phi product family. KNL has generated significant interest in the data analysis and machine learning communities because its new many-core architecture targets both of these workloads. The KNL many-core vector processor design enables it to exploit much higher levels of parallelism. At the Lincoln Laboratory Supercomputing Center (LLSC), the majority of users are running data analysis applications such as MATLAB and Octave. More recently, machine learning applications, such as the UC Berkeley Caffe deep learning framework, have become increasingly important to LLSC users. Thus, the performance of these applications on KNL systems is of high interest to LLSC users and the broader data analysis and machine learning communities. Our data analysis benchmarks of these application on the Intel KNL processor indicate that single-core double-precision generalized matrix multiply (DGEMM) performance on KNL systems has improved by ~3.5x compared to prior Intel Xeon technologies. Our data analysis applications also achieved ~60% of the theoretical peak performance. Also a performance comparison of a machine learning application, Caffe, between the two different Intel CPUs, Xeon E5 v3 and Xeon Phi 7210, demonstrated a 2.7x improvement on a KNL node.**

## I. INTRODUCTION

The MIT Lincoln Laboratory Supercomputing Center (LLSC) provides supercomputing capabilities to over 1000 users at MIT [1, 2]. Increasingly, these users require capabilities that are found in four ecosystems: supercomputing, databases, enterprise computing, and big data [3]. Each of these ecosystems has its own technologies with its own advantages (see [4]).

The LLSC has developed the MIT SuperCloud environment that allows all four ecosystems to run on the same hardware without sacrificing performance [3]. The MIT SuperCloud has spurred the development of a number of cross-ecosystem innovations in high performance databases [28, 29], database management [30], database federation [31, 32, 33], data analytics [34], data protection [35], and system monitoring [36, 37].

The LLSC continues to benchmark new computing technologies in order to deliver the best performance to its diverse user base.

The Intel Knights Landing (KNL) processor is an intriguing architecture because it can be self-hosting; i.e., it does not need to be hosted by a standard processor as current GPUs, or its predecessor, KNC (Knights Corner), processors do. In addition, users can run their existing codes on the KNL systems without any modifications. Since many LLSC users' data analysis applications are written for MATLAB and Octave, it is important to investigate how much performance improvement KNL systems can achieve as compared to LLSC's current Intel Haswell (Xeon E5-2683 V3 CPU @ 2.0 GHz) based systems.

There are a variety of reports about performance results on the KNL processors. Surmin et al. [15] reported performance results for the particle-in-cell plasma simulation code PICADOR [16] on KNL. They reported that a straightforward rebuilding of the code yielded a 2.43x speedup compared to the previous KNC generation. They also reported that further code optimization resulted in an additional 1.89x speedup. They mentioned that the optimization work was beneficial not only for KNL but also for high-end Intel Xeon CPUs and KNC. Their optimized version achieves 100 GFLOPS double-precision performance on a KNL processor with the speedups of 2.35x compared to a 14-core Haswell CPU and 3.47x compared to a 61-core KNC.

Researchers at the University of Edinburgh Edinburgh Parallel Computing Center (EPCC) [17] have reported on their benchmarks with KNL. These benchmarks used the KNL 7210 running at 1.30 GHz, with 64 cores, 16 GB multichannel dynamic random access memory (MCDRAM), and access to 96 GB DDR4 RAM running at 2133 MT/s. Their benchmarks consisted of three applications, COSA (CFD simulation code), GS2 (Gyro-kinetic simulation code), and CASTEP (density functional theory materials simulation code). These are all FORTRAN codes parallelized with MPI. The results are compared among KNC, KNL, Intel Ivy Bridge, and Intel Broadwell CPU systems. The reported performance is comparable between the KNL and Ivy Bridge system, with performance ~20% to 50% slower on the KNL, but the KNL achieves much faster performance than obtained with the KNC system. However, the KNL is lagging in performance when compared to the Broadwell system. The performance on the Broadwell is ~1.3x to 3.8x faster than that on the KNL. Interestingly, the performance impact of high-bandwidth memory (MCDRAM) varies among the applications. By running the applications on MCDRAM, the performance gain was varied significantly from no impact (CASTEP) to a

This material is based upon work supported by the Assistant Secretary of Defense for Research and Engineering under Air Force Contract No. FA8721-05-C-0002 and/or FA8702-15-D-0001. Any opinions, findings, conclusions or recommendations expressed in this material are those of the author(s) and do not necessarily reflect the views of the Assistant Secretary of Defense for Research and Engineering.

modest (COSA) 1.25x to a significant (GS2) 1.8x improvement. It was noted that while MCDRAM provides higher bandwidth than standard main memory, it has also slightly higher latency due to the extra logic needed in addressing the different banks of memory in MCDRAM. So MCDRAM will not improve performance on applications whose performance is dominated by memory latency.

Our benchmarking work is focused on a single-core and single-node performance on a KNL system, which is representative of the typical computing scenarios encountered by LLSC users. Two benchmark applications are considered that are representative of LLSC data analysis and machine learning workloads. First is a matrix multiplication (DGEMM) application written in the MATLAB M language and run using MATLAB and Octave. Second is a machine learning application called Caffe [18], which was developed by the Berkeley Artificial Intelligence Research (BAIR) Lab [19]. In order to benchmark Caffe performance, a public domain example is used, and its performance is compared between the Intel Haswell- and KNL-based LLSC systems.

## II. LLSC SYSTEMS

LLSC cluster systems comprise many different types of compute nodes that are manufactured by different vendors. This study focused on Intel processor systems. In particular, two different Intel nodes, one with dual-socket Haswell (Intel Xeon E5-2683 V3 @ 2.0 GHz) processors and another single-socket KNL (Intel Xeon Phi KNL 7210 @ 1.3 GHz). Each Haswell processor has 14 cores and can run two threads per core with the Intel hyper-threading technology. Each Haswell node has 256 GB of memory. The KNL 7210 processor has 64 cores and four threads per core. Each KNL server node has 192 GB of main memory. In addition, a stand-alone KNL workstation with a single KNL 7210, which has 96 GB of memory, was also benchmarked. Both the KNL workstation and servers have 16 GB of high-bandwidth memory (MCDRAM).

MCDRAM can be set up in three different modes on the KNL: flat, cache and hybrid [20]. On LLSC KNL nodes, MCDRAM is configured as flat memory where MCDRAM and main memory are two separate memory spaces. In addition, the KNL mesh-interconnect has BIOS configuration options for three different cluster modes: all-to-all, quadrant, and sub-NUMA clustering [20]. LLSC KNL systems are configured to use the quadrant clustering mode, which divides the processor into four virtual quadrants. This arrangement provides lower latency and higher bandwidth than the all-to-all mode. This mode is recommended for applications that treat KNL as a symmetric multiprocessor (SMP).

## III. BENCHMARK RESULTS

The MATLAB and Octave DGEMM benchmarks were first performed on the KNL workstation because we wanted to evaluate the performance ahead of the purchase of the KNL cluster systems. With the purchase of the KNL cluster system, which consists of 41,472 cores, we had run the standard LINPACK benchmark. On the biannual Top 500 supercomputer list, which was published in November 2016, the LLSC KNL cluster system was ranked the most powerful supercomputer in New England and the third most powerful at a United States university [21]. The DGEMM benchmark was performed on a KNL server node and the results were compared with those obtained earlier from the KNL workstation.

Since the KNL cluster system is specifically focused on enabling new research in machine learning, advanced physical devices, and autonomous systems, we have also worked on benchmarking one of the popular machine learning applications, Caffe. We compared the results between the Intel Haswell and KNL nodes and demonstrated how much speedup the KNL server can achieve. We expect that the KNL performance results may improve as the underlying software is further optimized for the KNL architecture.

### A. *Data Analysis Application Single-Core Performance*

The MATLAB/Octave DGEMM benchmark is a simple double-precision matrix-matrix multiplication code with varying square matrix sizes. Figure 1 presents the single-core MATLAB and Octave DGEMM performance. The vertical axis represents the floating-point performance in MFLOP/s. Since MATLAB enables using an external math library, the results with MATLAB have been obtained by using both its own and Intel MKL version 110303 math libraries.

The Octave binary built with the OpenBLAS version 0.2.18 library shows the worst performance in Figure 1. This result indicates that the OpenBLAS version 0.2.18 library is not fully optimized for the KNL architecture and is not recompiled on KNL since we used the system-provided library. In addition, the performance with MATLAB's own math library shows a little better performance compared to that of the Octave built with the OpenBLAS library, but its performance is still quite behind that of the Intel MKL math library, which is tuned for the KNL architecture.

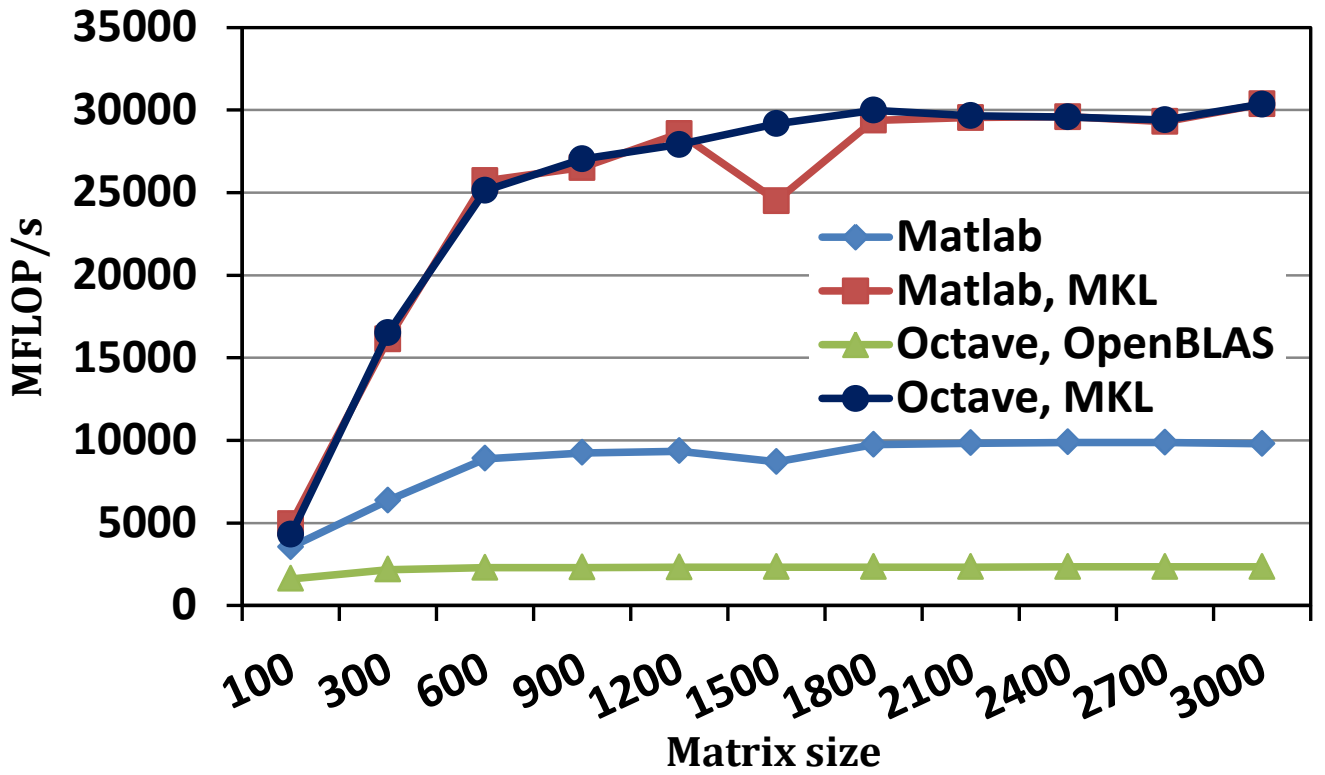

Fig. 1. A single-core DGEMM performance with various matrix sizes for MATLAB and Octave using different math libraries on a KNL 7210 workstation.

Both Octave built with the Intel MKL version 110303 library and Matlab using the same Intel MKL math library have shown similar high performance. The MKL math library offers much better performance than the other math libraries tested in this experiment. It was not clear why MATLAB DGEMM performance shows a sudden drop at the matrix size of 1500, but this behavior was consistent throughout the rest of the results.

When the KNL cluster system was installed at LLSC, the Intel compiler was upgraded along with the MKL version 20170000 math library. Figure 2 compares the DGEMM performance obtained on a KNL server node using the Intel MKL version 20170000 library with the earlier results obtained on the KNL workstation with the MKL version 110303 library. Overall, the new MKL library provides better DGEMM performance throughout the whole range of the matrix sizes. At the matrix size of 3000, the performance was improved 13%. The single-core DGEMM performance with MATLAB on a KNL node improves ~3.5x using KNL version 20170000 math library, compared to the performance obtained using MATLAB's default math library. It is interesting to note that Octave performed a little better than MATLAB on the KNL server node although they are using the same MKL library.

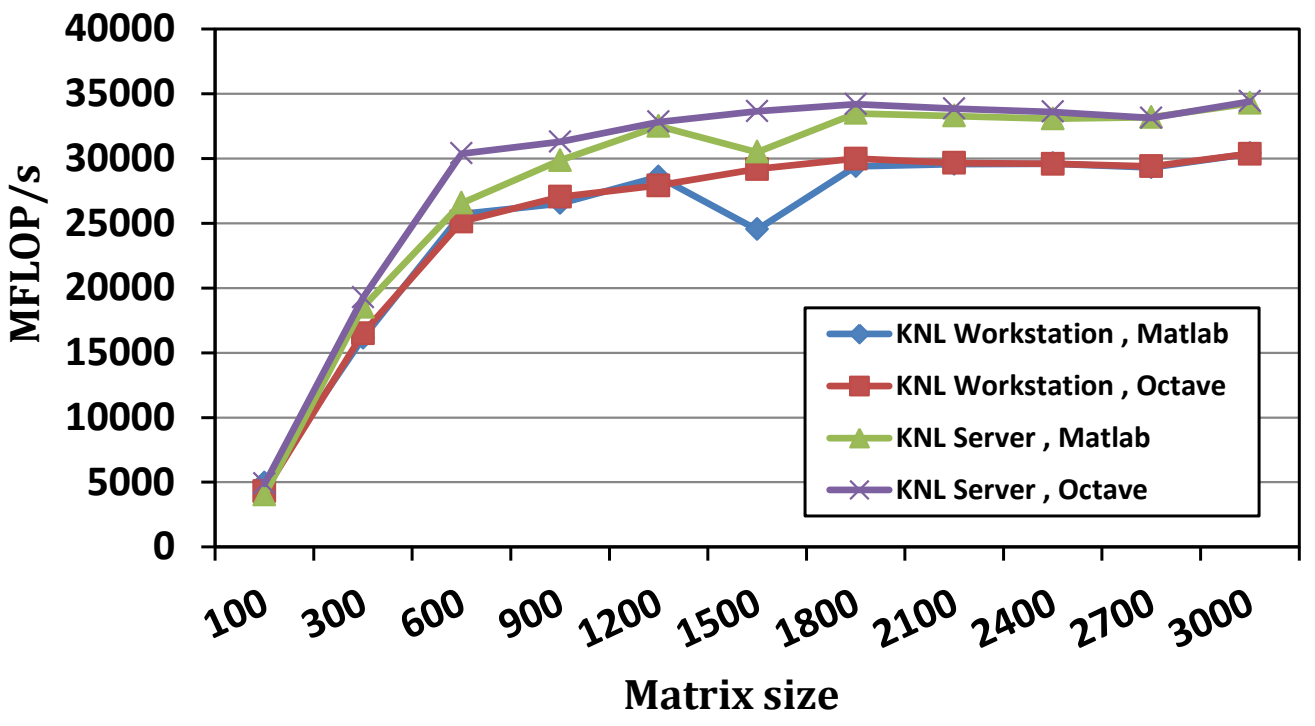

Fig. 2. Comparison of a single-core DGEMM performance for MATLAB and Octave with various matrix sizes on a KNL workstation and a KNL server node using two different MKL versions, 110303 and 20170000, respectively.

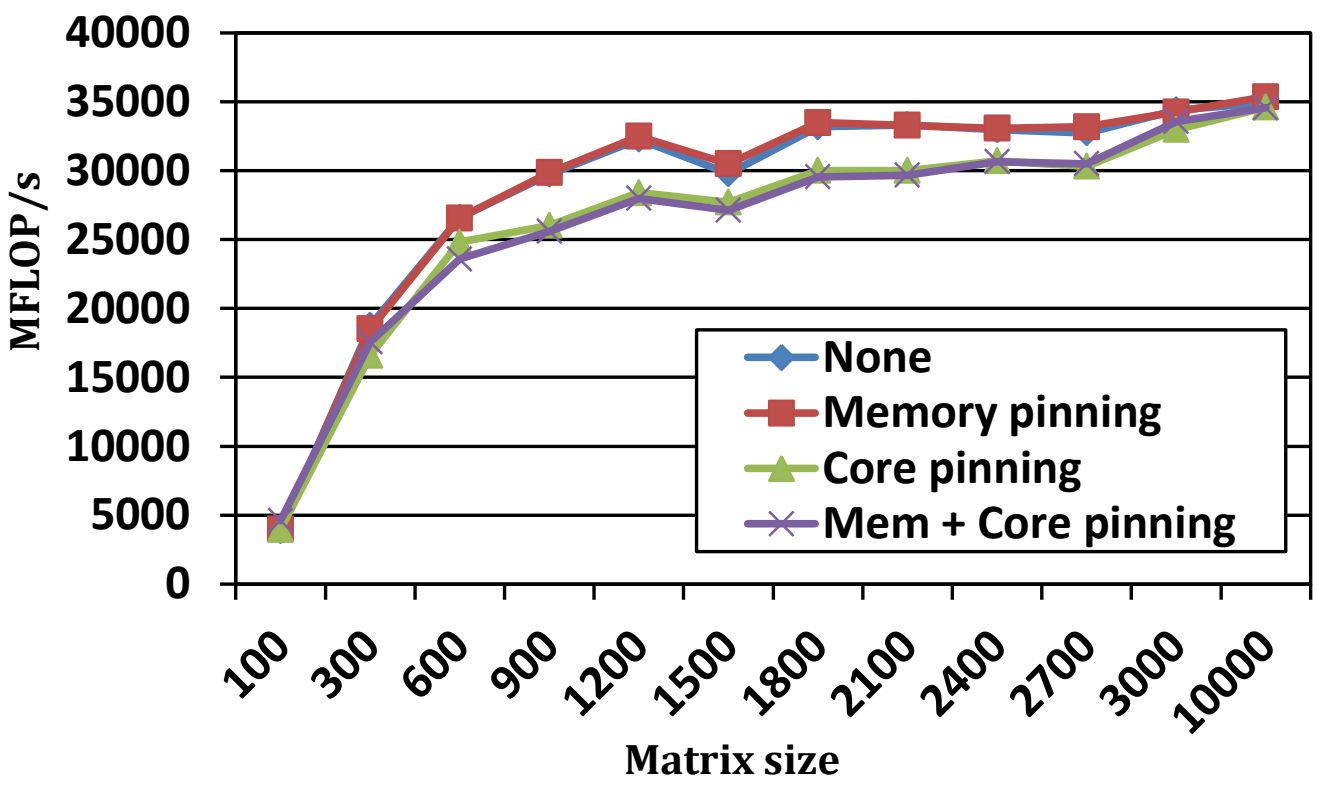

Fig. 3. A single-core DGEMM performance for MATLAB with various matrix sizes using different NUMA strategies.

Figure 3 shows how the single-core DGEMM performance varies with different NUMA strategies. As shown in Figure 3, the best performance was obtained by pinning the process on the MCDRAM memory. Interestingly, the performance with no NUMA control produced similar performance to that achieved by pinning the process to the MCDRAM memory. It is believed that because the DGEMM code is CPU intensive, the memory speed is not as important. Furthermore, adding core-pinning CPU 0 to the memory pinning caused an adverse effect. Finally, core-pinning alone did not produce better results.

Figure 4 compares the performance difference of a single-core DGEMM benchmark between the Intel Haswell and KNL processors. It shows the results for both MATLAB and Octave using the Intel MKL version 20170000 math library. The Haswell processor performs slightly better for both MATLAB and Octave, which is about 15% and 10% faster with the matrix size of 3000, respectively. Although the Haswell processor speed (2.0 GHz) is about 54% faster than that of the KNL processor (1.3 GHz), the performance gap is not as big as its speed difference because the new KNL processor AVX512 vector processing engines are two times wider than the AVX2 engines in the Haswell processor.

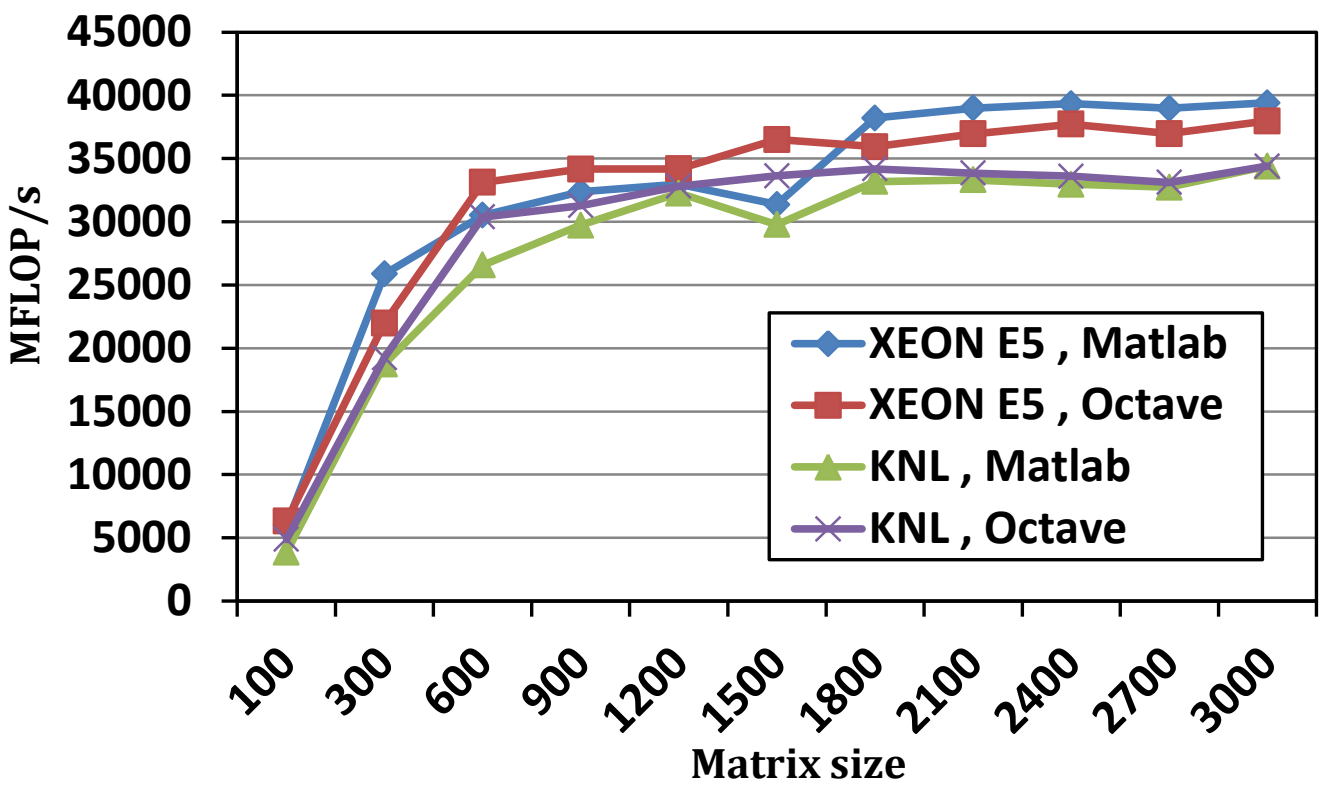

Fig. 4. DGEMM performance difference of Intel Haswell versus KNL 7210 processors for MATLAB and Octave.

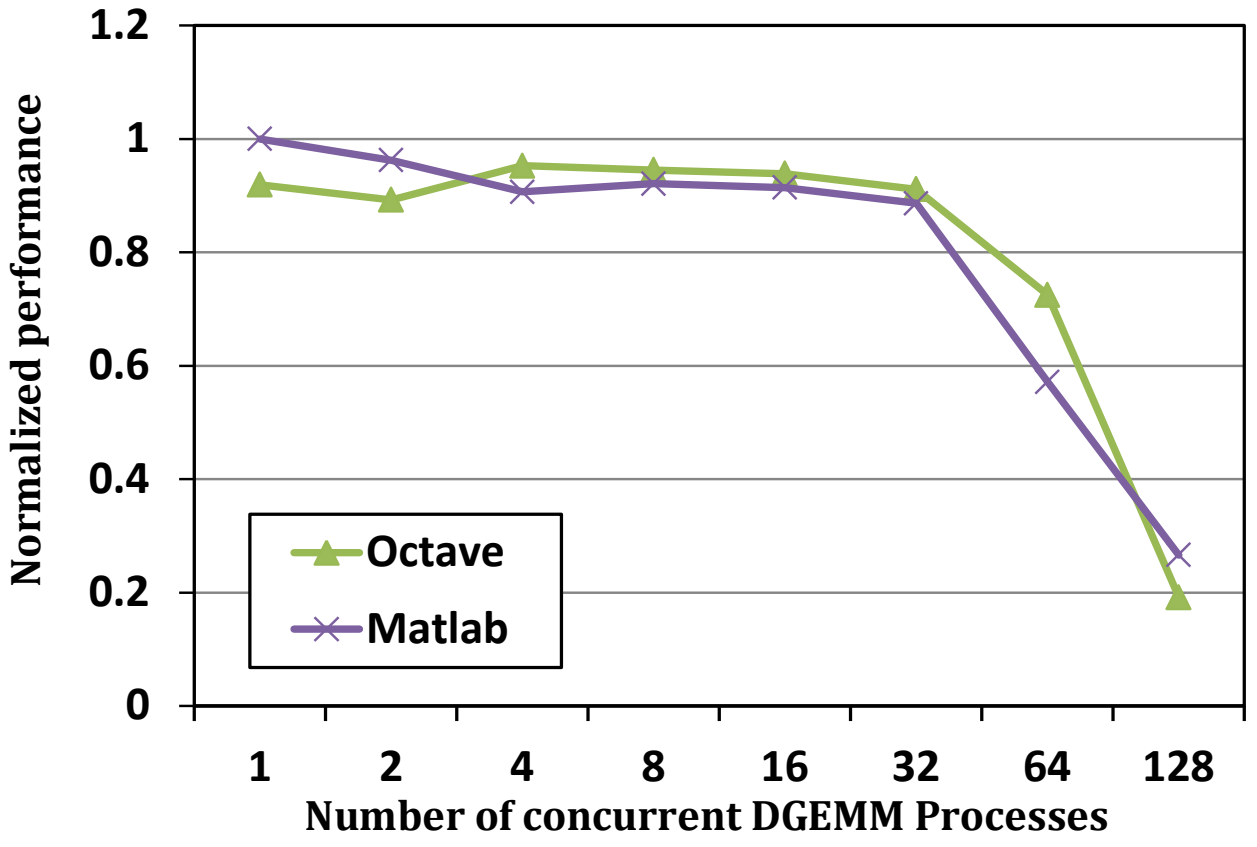

Fig. 5. Normalized average single-core DGEMM performance with a fixed matrix size of 3000 while running multiple instances of DGEMM processes.

### B. *DATA ANALYSIS APPLICATION MULTI-CORE PERFORMANCE*

A primary feature of the KNL is the large number of cores available on each processor. Thus, it is important to measure the throughput performance on a KNL node when multiple processes of a single-thread DGEMM benchmark are running concurrently on the same node. For this experiment, the matrix size was set to 3000 and run 10 times to average the performance. The DGEMM processes were launched using our internally developed parallel library [22, 23, 24]. In order to make sure that all the DGEMM calculations are started at the same time, a barrier is called among all the concurrent MATLAB processes at the beginning of DGEMM calculation.

The result is saved in the memory of each process, and all the results are collected by the lead process at the end. As shown in Figure 5, the horizontal axis represents the number of concurrent DGEMM processes, and the vertical axis is the normalized DGEMM performance by the single core Matlab performance. When the number of concurrent DGEMM processes increases, the average DGEMM performance per process stays at relatively the same level up to 32 concurrent processes but drops significantly afterwards.

The normalized throughput performance on a single node when it is running various numbers of concurrent DGEMM processes is presented in Figure 6. Both MATLAB and Octave behave similarly up to 32 concurrent processes, but significantly different throughput performance is observed when the KNL node is fully populated with 64 concurrent processes or overloaded. Octave performed better than MATLAB with 64 concurrent processes because Octave occupies significantly less memory than MATLAB. The maximum throughput performance on a KNL node was achieved with Octave and is ~60% of its theoretical peak performance. As shown in the figure, when the KNL node is overloaded with more DGEMM processes than the number of physical cores, the total throughput performance was significantly lower than its peak performance although its drop is less significant with MATLAB.

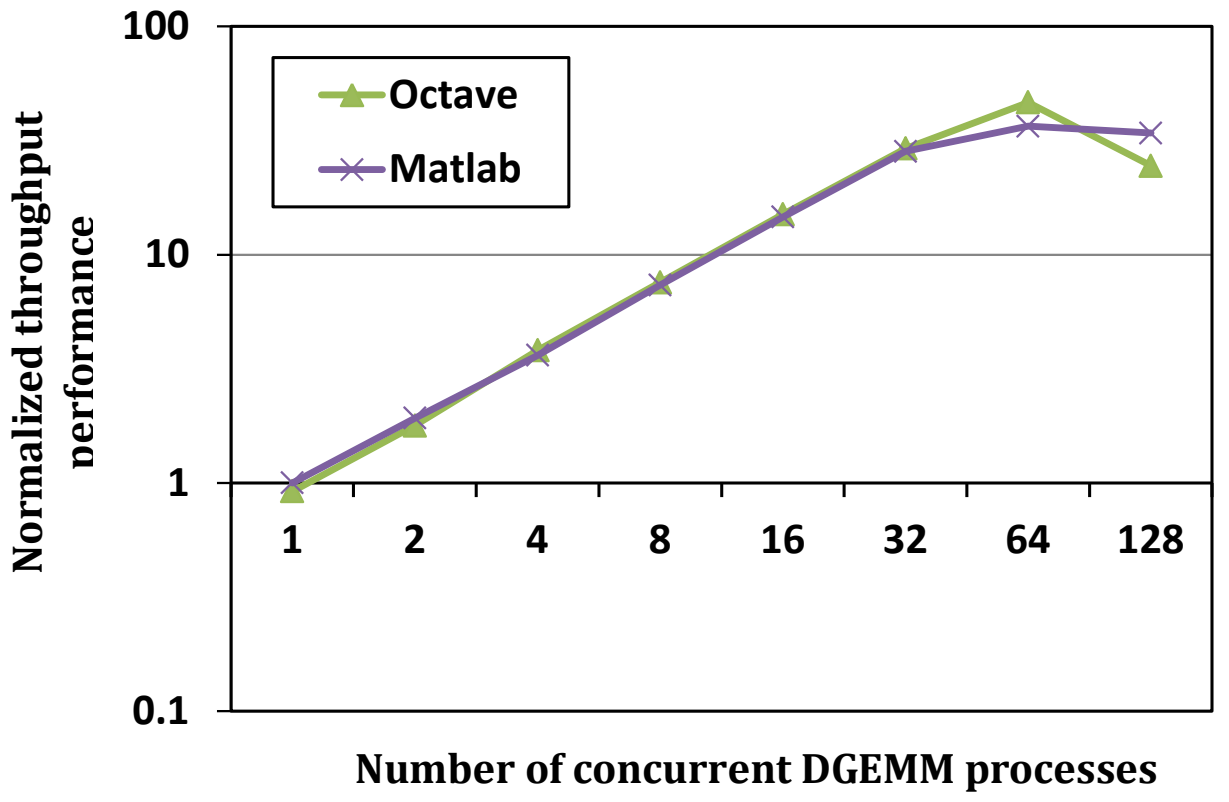

Fig. 6. Total throughput performance of various numbers of DGEMM processes with a fixed matrix size at 3000 by 3000.

### C. *MACHINE LEARNING APPLICATION*

The rise of effective machine learning has resulted in the development of many deep learning frameworks and libraries, such as Caffe [18], TensorFlow [25], Torch [26], scikit-learn [27], Theano [28], and many more [29]. There is great interest in measuring the performance of machine learning software on KNL. Although it would be ideal to investigate all of the software mentioned above, we have chosen Caffe as a representative example. Caffe is a deep learning framework developed by Berkeley AI Research Lab [19] and has implemented several published CNN (Convolution Neural Network) models and tutorials such as AlexNet [30, 31], LeNet [32, 33], GoogleNet [34, 35], and an example tutorial for ImageNet data [36]. For this work, we selected an example that finetunes CaffeNet for style recognition on "Flickr style" data [37, 38] with using the default single-precision arithmetic. The term, finetuning, means that the new model takes an existing trained model, CaffeNet in this case, adapts the architecture, and resumes training from the already learned model weights [38]. For these measurements, we have used 6421 images downloaded from the Flickr website using the script provided in the tutorial [38].

Since we are interested in how Caffe behaves on both Intel Haswell and KNL processors, we first looked at its performance trends based on different numbers of batch sizes when the model is being trained on the images. The batch size is the number of images sampled from the training data. The sampled images are used to approximate the exact gradient of the parameters with respect to the training data, and the changes of the parameters are made in the direction of the gradient. If the size of the training data is large, increasing the batch size may result in more accurate estimation of the gradient, and in turn change the results of the training process.

Figures 7 and 8 show the number of images that are processed per second when the batch sizes are varied on Intel Haswell and KNL systems, respectively. The average forward pass time is the time it takes to compute the output given the input for inference, where Caffe computes the function represented by the model at each layer and composes them all. The average backward pass time is the time it takes to compute the gradient given the loss for learning [39]. The average forward-backward time includes the update time as well. The performance has been measured by running 1000 iterations of Caffe when it is fully utilizing the node; in other words, 28 threads are used for the Haswell node and 64 threads are used for the KNL node. Increasing the batch size on the Haswell server significantly improves the performance of the forward pass, and its peak performance is reached at the batch size of 512.

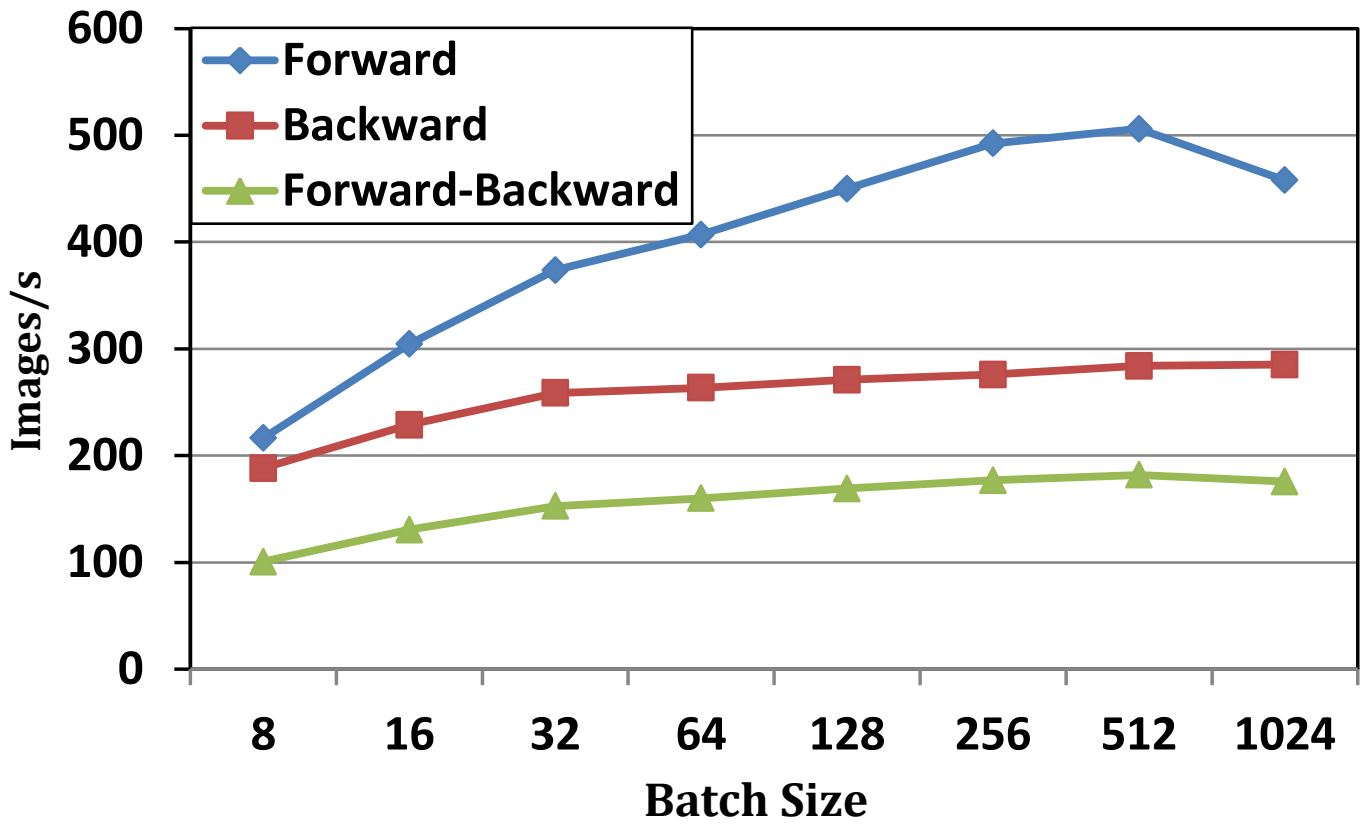

Fig. 7. Number of images processed per second with varying the batch sizes on an Intel Haswell server.

On the KNL server as shown in Figure 8, the forward pass performance increases in a similar fashion as seen with the Haswell server. However, the performances of the backward and forward-backward passes are increasing at a much faster rate on the KNL server than those observed on the Haswell server as increasing the batch sizes. In other words, Caffe performs better on both forward and backward passes and results in a higher processing rate on the KNL server. It is also noted that because of the memory size of MCDRAM, we can increase the batch size up to 1088, where its performance becomes worse than the performance at the batch size of 1024. It is noted that the performance results presented on the KNL server are obtained while the process is pinned in MCDRAM. It is observed that the performance results of Caffe by pinning its process on MCDRAM is about 24 ~ 29% faster than the results obtained without the NUMA control for the memory binding.

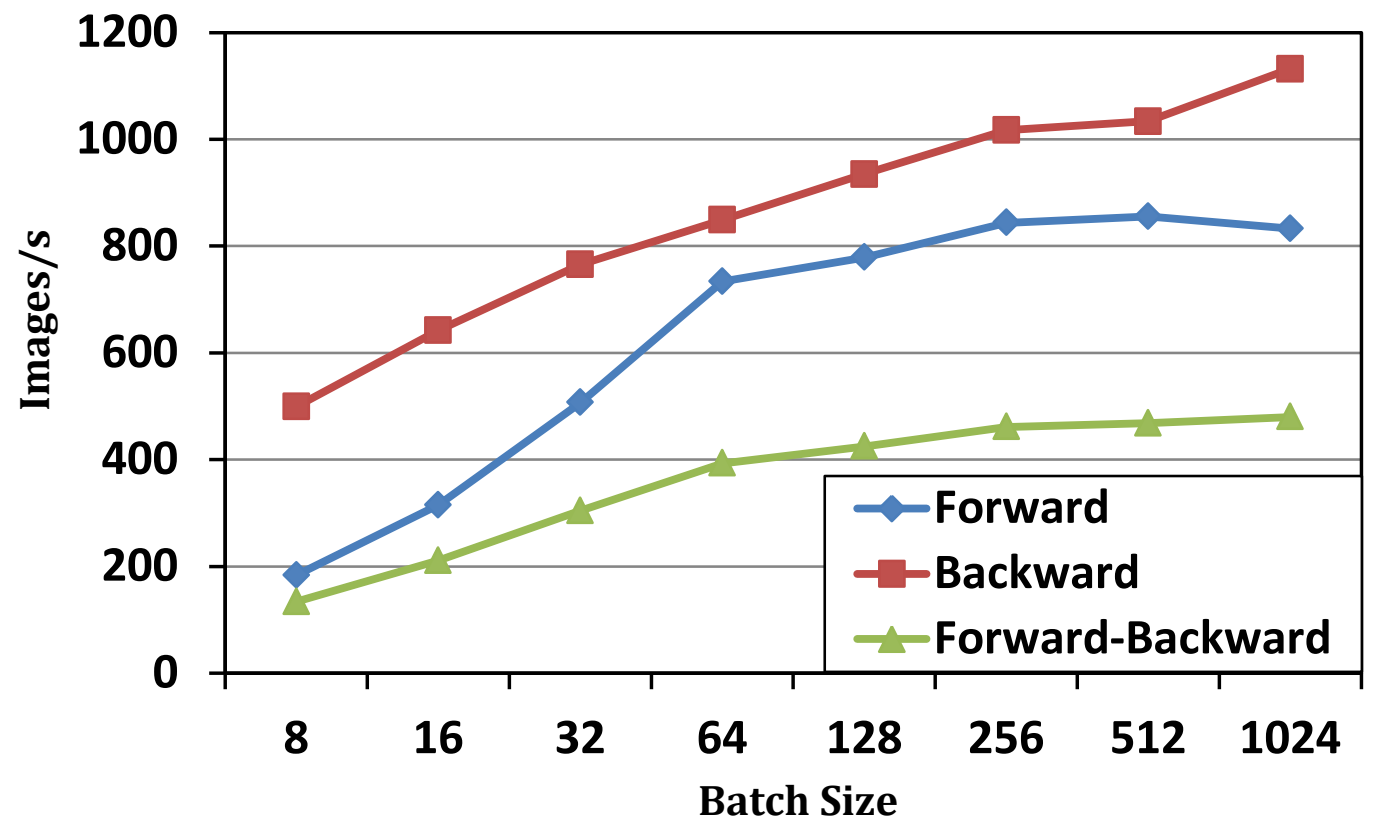

Fig. 8. Number of images processed per second with varying the batch sizes on an Intel KNL server.

As shown in Figure 9, the overall performance improvement on the Intel KNL server is as much as 2.7 times better than that observed on the Intel Haswell server. The performance gain on the KNL server is rapidly increasing at smaller batch sizes but flattens out as the batch size increases beyond 32. Since the Haswell node shows diminishing performance while the KNL node is still increasing in performance at the batch size of 1024, the performance ratio between the two nodes increases at the batch size of 1024.

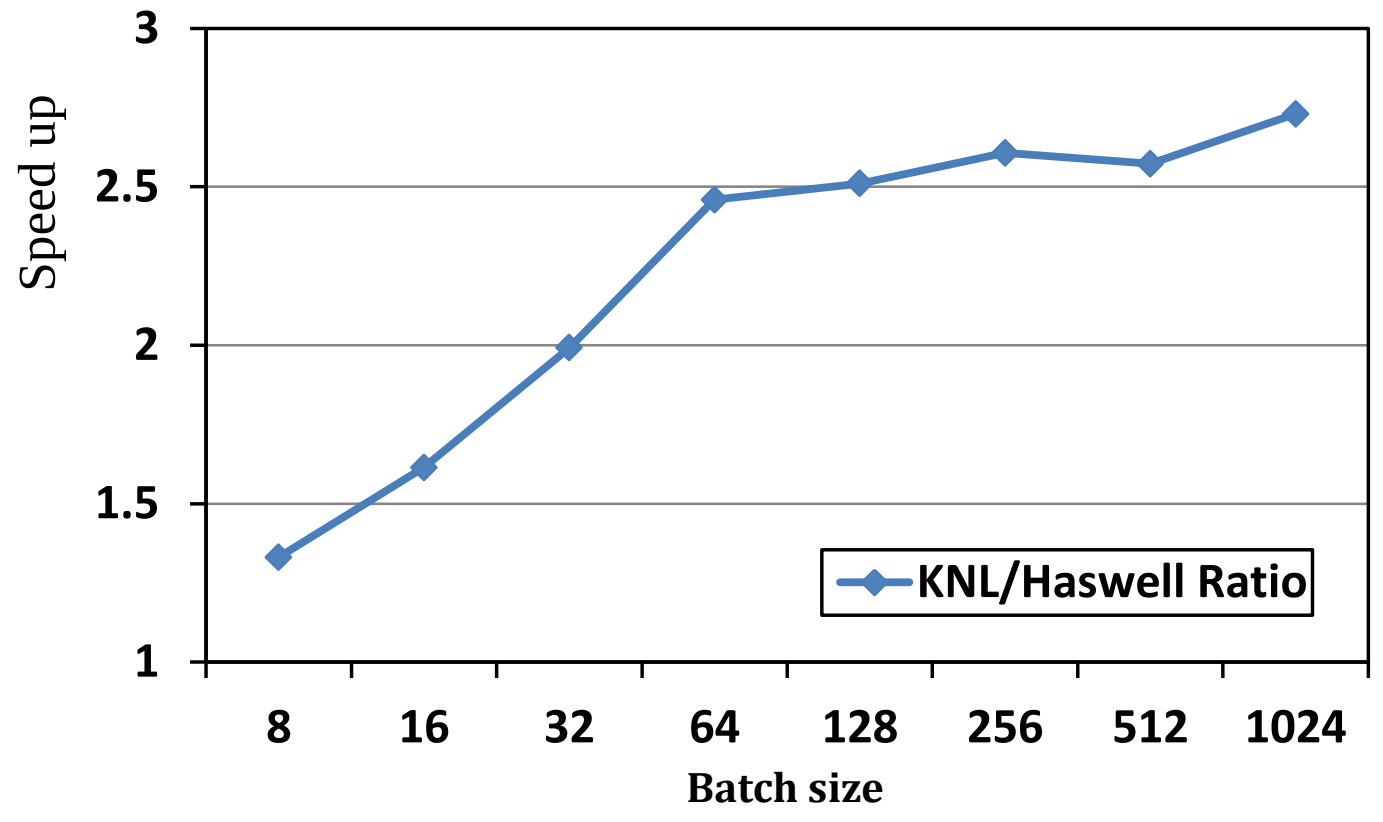

Fig. 9. The speedup ratio in terms of the processed image rate with respect to the batch sizes between the results obtained on Intel Haswell and KNL servers.

## IV. SUMMARY

We have presented the performance results of the KNL processor using a data analysis application (DGEMM) written in MATLAB M language and a machine learning application using the Caffe deep learning framework. These applications are selected because they represent the core functionality of the current and future usage scenarios at the LLSC.

As demonstrated in the DGEMM benchmark results, to achieve the best performance on the KNL node, it is important to use the math library that is optimized for the KNL architecture.

NUMA control on the core-binding with the DGEMM benchmark can have an adverse impact on its performance while the memory binding to MCDRAM has little impact on the DGEMM performance. It is believed this observation is because the DGEMM code is CPU intensive and the memory speed is not as important.

From our benchmark results, we have observed that the single-core DGEMM performance on KNL systems has improved about 3.5 times with the KNL-optimized MKL math library as compared to MATLAB's own math library. The throughput performance achieved on a KNL node was ~60% of its theoretical peak performance.

We have also investigated how the batch size affects the performance of Caffe with an example from a tutorial using image data downloaded from Flickr. We have observed that the best performance is achieved for the Haswell and KNL nodes at the batch size of 512 and 1024, respectively. The KNL node achieves ~2.7x better performance as compared to the Haswell node. Recently Caffe2 [40] has been released to the public and we are looking forward to benchmarking this latest release.